\documentclass[twocolumn,floatfix,superscriptaddress,longbibliography,nofootinbib]{revtex4-2}
\RequirePackage[sort&compress]{natbib}
\usepackage[english]{babel}
\usepackage{color}
\usepackage{subcaption}
\usepackage{booktabs}
\usepackage{float}

\usepackage{ifthen}
\usepackage{siunitx}
\usepackage{tikz}
\usetikzlibrary{hobby} 
\usetikzlibrary{arrows.meta} 
\tikzset{>={Latex[length=4,width=4]}} 
\usetikzlibrary{calc,intersections,decorations.markings}
\usepackage{siunitx}
\usepackage{xcolor} 

\colorlet{mylightblue}{blue!20}
\colorlet{myblue}{blue!50!black}
\colorlet{mydarkblue}{blue!30!black}
\colorlet{mylightred}{red!10}
\colorlet{myred}{red!50!black}
\colorlet{mydarkred}{red!60!black}
\colorlet{mydarkgreen}{green!30!black}

\tikzset{
  midarr/.style={decoration={markings,mark=at position #1 with {\arrow{stealth}}},postaction={decorate}},
  midarr/.default=0.5
}

\usepackage{amssymb,amsfonts,amsmath}
\usepackage{array}
\usepackage{bm}
\usepackage{xcolor}
\usepackage{comment}

\usepackage[colorlinks = true,
            linkcolor = blue,
            urlcolor  = blue,
            citecolor = blue,
            anchorcolor = blue]{hyperref}

\usepackage{dcolumn}
\usepackage{bm}
\usepackage{float}

\renewcommand{\[}{\left[}

\usepackage{xcolor}
\definecolor{RoyalBlue}{HTML}{4169e1}
\definecolor{ForestGreen}{HTML}{228b22}

\usepackage{setspace}
\usepackage{comment}

\usepackage{todonotes}

\begin{document}


\title{FKPP fronts in quenched random media}

\author{Ulysse Marquis}
\email{ulyssepierre.marquis@unitn.it}
\affiliation{Fondazione Bruno Kessler, Via Sommarive 18, 38123 Povo (TN), Italy}
\affiliation{Department of Mathematics, University of Trento, Via Sommarive 14, 38123 Povo (TN), Italy}

\author{Henri Berestycki}
\affiliation{Department of Mathematics, University of Maryland, College Park, MD, USA}
\affiliation{Centre d'Analyse et de Math\'ematique Sociales, Ecole des Hautes Etudes en Sciences Sociales, Paris, France}
\affiliation{Institute of Advanced Study, Hong Kong University of Science and Technology, Hong Kong}

\author{Marc Barthelemy}
\affiliation{Universit\'e Paris-Saclay, CNRS, CEA, Institut de Physique Th\'eorique, 91191, Gif-sur-Yvette, France}
\affiliation{Centre d'Analyse et de Math\'ematique Sociales, Ecole des Hautes Etudes en Sciences Sociales, Paris, France}
\affiliation{Complexity Science Hub,
Metternichgasse 8, 1030, Vienna, Austria}

\begin{abstract}
We study numerically the evolution of one-dimensional FKPP fronts initiated from steep initial conditions in the presence of a quenched random growth rate. Compared to both the homogeneous case (with velocity $v_0$) and deterministic disorder, quenched randomness increases the average propagation speed. We show that the velocity shift relative to the homogeneous case scales linearly with the disorder variance $\sigma^2$, with a universal prefactor---independent of the specific distribution of the disorder---such that 
$v = v_0 + a \sigma^2$, with $a \approx 0.02432 \pm 0.00002$. Moreover, the front position exhibits diffusive fluctuations across disorder realizations. The corresponding effective diffusion coefficient scales quadratically with $\sigma$, 
$D = \frac{b^2 \sigma^2}{2}$, with $b \approx 0.223 \pm 0.002$. These results suggest a universal statistical response of FKPP fronts to quenched heterogeneity.
    
\end{abstract}

\maketitle

\section{Introduction} 

Reaction--diffusion equations are a cornerstone of the theoretical description of growth, invasion, and spreading phenomena across physics, biology, and the social sciences. Among them, the Fisher--Kolmogorov--Petrovskii--Piskunov (FKPP) equation~\cite{Fisher:1937, KPP} occupies a paradigmatic position. Originally introduced in the context of population genetics, the F-KPP equation describes the competition between local growth and spatial diffusion, leading to the emergence of traveling fronts with a well-defined asymptotic velocity~\cite{KPP, aw2, VANSAARLOOS_2003}.

In its simplest one-dimensional form, the FKPP equation reads
\begin{equation}
\partial_t u(x,t) = D\,\partial_{xx} u(x,t) + K_0\,u(x,t)\bigl(1-u(x,t)\bigr),
\label{eq:kpp}
\end{equation}
where $u(x,t)$ is a normalized density field, $D$ is the diffusion coefficient, and $K_0>0$ is the growth rate. For compactly supported or sufficiently steep initial conditions, the solutions of Eq.~\eqref{eq:kpp} travel with asymptotic speed
\begin{equation}
v_0 = 2\sqrt{D K_0},
\end{equation}
a result that can be derived from the linearized dynamics at the leading edge of the front~\cite{KPP, aw2, VANSAARLOOS_2003}.

This property makes FKPP fronts a canonical example of \emph{pulled fronts}, whose asymptotic behavior is selected by the low-density region ahead of the bulk.


Beyond its original motivations, the FKPP equation has found applications in a wide range of physical and biological contexts. Biological invasions~\cite{ske,Shigesada_1997} provide one of the most compelling testbeds for such systems. To leading order, the spatial spread of a species can be modeled as a reaction--diffusion process combining individual dispersal (diffusion) and population growth (reaction). The effects of environmental heterogeneities~\cite{Shigesada_1997}, habitat fragmentation~\cite{Berestycki_2005,Berestycki_2005_2,roques_2007}, and obstacles~\cite{Berestycki_2009} on the existence, speed, and shape of traveling waves have attracted sustained attention. 

Similar reaction--diffusion frameworks arise in a variety of contexts, including flame propagation~\cite{Zeldovich1938}, epidemic spreading, population biology~\cite{Murray_2003}, and chemical kinetics. More recently, such ideas have been applied to urban systems. An empirical study of the spatial growth of cities~\cite{Marquis_2025} analyzed the dynamics of the interface of the giant urban component and reported a lack of universality in the standard growth and dynamic exponents, pointing to limitations of classical scaling approaches. Earlier works had already proposed describing urban expansion through partial differential equations inspired by reaction--diffusion processes~\cite{friesen2019,whiteley2022,capeltimms2024}; see also the reviews~\cite{marquis_review,Barthelemy_2025}. Within this perspective, neighborhood interactions generate diffusive coupling, while land scarcity, congestion, and resource constraints introduce saturation effects, naturally leading—at a coarse-grained level—to FKPP-type models for urban density. However, real cities evolve in strongly heterogeneous environments shaped by geography, infrastructure, resource distribution, and policy constraints. This has motivated the study of generalized FKPP equations with spatially varying coefficients,
\begin{equation}
\partial_t u(x,t) = D \, \partial_{xx} u(x,t) + K(x)\,u(x,t)\bigl(1-u(x,t)\bigr),
\label{eq:fkpp_hetero}
\end{equation}
where the growth rate $K(x)$ encodes spatial heterogeneity~\cite{bhn-jams,bhna1,whiteley2022,capeltimms2024,marquis_review}.

FKPP equations with various forms of annealed disorder have been investigated in, e.g.,~\cite{mueller,armero1996,brunet1999,Tripathy_2001,Brunet:2001,Rocco_2001,Rocco_2002,brunet2004,conlon2005,Brunet_2006,Brunet_2006_2,Nolen2011,mueller2019}; see also Chapter~7 of~\cite{VANSAARLOOS_2003} for a review. Of particular interest here is the case in which $K(x)$ is random but time-independent (i.e., quenched disorder), reflecting frozen spatial heterogeneity rather than temporal fluctuations. We also note that the randomized FKPP equation is closely related to the problem of the maximum of branching Brownian motion in random potentials~\cite{cerny2024}. In such settings, spatial variability can strongly influence front propagation. Understanding how quenched randomness in the growth rate affects spreading velocities and interface fluctuations is therefore essential for assessing the robustness of FKPP-based descriptions in heterogeneous growth processes, including urban expansion.

From a physical perspective, disorder in the growth rate fundamentally alters the mechanism of front propagation. Since FKPP fronts are pulled, their asymptotic velocity is controlled by the dynamics in low-density regions at the leading edge of the front. In heterogeneous environments, these regions sample the spatial distribution of the local growth rate $K(x)$, making the spreading process highly sensitive to fluctuations of this field. As a result, the front velocity is no longer determined solely by the mean growth rate, but instead depends on the disorder. 

In one dimension in particular, even weak quenched heterogeneity can lead to systematic deviations from the homogeneous FKPP velocity. These deviations were first investigated in~\cite{Gartner_1979}, where the spreading speed was shown to be determined by the principal eigenvalue of an associated elliptic operator in periodic media. In contrast, in random stationary ergodic media, the spreading speed is linked to the statistical properties of the heterogeneity; see Theorem~2 of~\cite{Gartner_1979}. Building on these ideas,~\cite{Berestycki_2005, Berestycki_2005_2} provided a 
more general treatment for large classes of equations both in the periodic framework and hetereogeneous medium case. Nadin 
and coauthors~\cite{Nadin_2010,Nadin_2015,Hamel_2011} investigated the effect of heterogeneity in the case of random ergodic media. 
Further works~\cite{Berestycki_2012, bna} continued this effort in the cases of random ergodic stationary, almost periodic media and general deterministic heterogeneous environments. Specifically,~\cite{Nadin_2015} found that adding null-average heterogeneity to the medium increases spreading velocity, and studied the effect of variations of the diffusion and reaction coefficients on the magnitude of the speed increase; in particular, comparing it to the homogeneous case. 

Understanding front propagation in quenched random fields provides a theoretical framework for growth in heterogeneous, frozen landscapes and clarifies how macroscopic spreading laws emerge from the interplay of diffusion, nonlinear saturation, and spatial disorder. In the following, we analyze how quenched heterogeneity affects FKPP fronts, with particular emphasis on the mechanisms that determine the asymptotic spreading speed and their implications for spatial growth processes.

\section{The model}

In this paper, we study the one-dimensional Fisher--KPP equation~\cite{Fisher:1937, KPP} in a quenched random reaction field
\begin{equation}\label{eq:fkpp}
    \partial_t \rho(x,t) = \partial_{xx} \rho(x,t) + \left[1 + 
    \eta(x)\right] \rho(x,t) \left[1 - \rho(x,t)\right],
\end{equation}
where $\eta(x)$ is a spatially random field, independently and identically distributed, with zero mean and delta covariance
\begin{align}
\langle 
\eta(x) \rangle = 0, \quad \langle \eta(x) \eta(x') \rangle = \sigma^2 \boldsymbol{\delta}(x - x'),
\end{align}
where $\sigma$ denotes the noise intensity, the brackets~$\langle \ldots \rangle$ represents averages over disorder realization and
where $\boldsymbol{\delta}(\cdot)$ denotes the Dirac delta distribution. The initial condition is chosen as a sharply decaying Gaussian profile,
\begin{align}
\rho(x,0) = \exp\left(-\frac{x^2}{2 s^2}\right).
\end{align}
Unless stated otherwise, the results presented in this paper were obtained with $s=2$. We verified that, at sufficiently long times, the asymptotic behavior is independent of the choice of $s$.\\

We define $x(t)$ as the right-most point such that $ \rho(x(t), t) = \tfrac{1}{2}$, and study its evolution through numerical integration of Eq.~\eqref{eq:fkpp}. We utilized an explicit integration method with~$\delta t = 10^{-3}$ and~$\delta x = 10^{-1}$, which respects the Courant–Friedrichs–Lewy condition.

We will consider several probability distributions for the quenched field $\eta(x)$:
\begin{itemize}
    \item[(i)] a uniform distribution on $[-\delta,\delta]$;
    \item[(ii)] a symmetric triangular distribution on $[-\delta,\delta]$;
    \item[(iii)] a symmetric bimodal distribution
    \begin{align}
        p(\eta)=\tfrac{1}{2}\,\boldsymbol{\delta}\!\left(
    \eta+\delta\right)+\tfrac{1}{2}\,\boldsymbol{\delta}\!\left(
\eta-\delta\right),
    \end{align}
    \item[(iv)] an asymmetric bimodal distribution
    \begin{align}
        p(\eta)=\lambda\,\boldsymbol{\delta}\!\left(\eta-
    \eta_-\right)+(1-\lambda)\,\boldsymbol{\delta}\!\left(\eta-\eta_+\right),
    \end{align}
    with $0<\lambda<1$ and parameters chosen such that the mean vanishes, $\lambda \eta_- + (1-\lambda)\eta_+ = 0$. 
    \item[(v)] an exponential distribution~$\eta = \sigma \left( \mathcal{E}-1 \right)$, with~$\mathcal{E}$ following a standard exponential of mean and variance~$1$.
\end{itemize}
All these distributions have zero mean and variances $V$ that can be computed explicitly. Note the diversity of probability laws considered: discrete and continuous, bounded or unbounded (on the right), peaked, bimodal, or uniform.

\section{Microscopic study}

Integrating numerically Eq.~\ref{eq:fkpp} yields well-defined fronts propagating into the unstable phase $\rho=0$, with fluctuating instantaneous speeds. 

\begin{figure}
    \centering
    \includegraphics[width=0.7\linewidth]{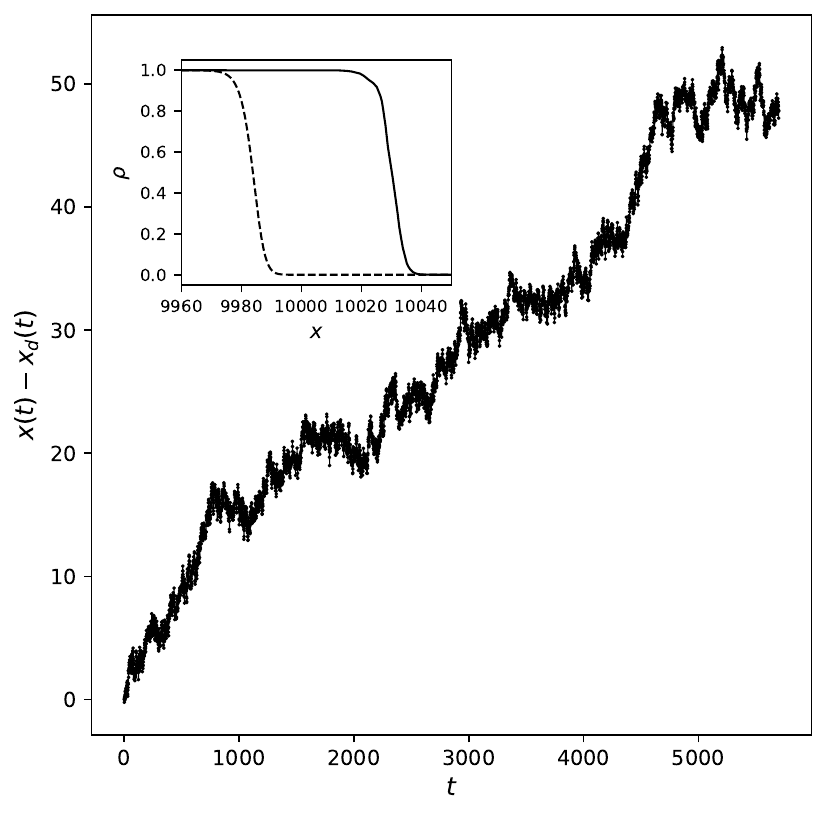}
    \caption{\textbf{Front progressions}. Main figure : difference of progression between a front propagating in an uniform disordered medium with~$\delta=1$ and a front in a homogeneous media~$x(t)-x_d(t)$. Inset : shape of the fronts (ordered medium : dashed line, disordered medium : solid line) at time $t=5000$.
    }
    \label{fig:micstudy}
\end{figure}

Figure~\ref{fig:micstudy} shows the positional shift between a front evolving in a uniformly disordered field with standard deviation $\sigma = 1/\sqrt{3}$ (case~(i), $\delta=1$) and its homogeneous counterpart. While its progression is noisy, it shows an upward trend suggesting an increase in asymptotic speed, of order given by~$\frac{x(t)-x_d(t)}{t} \approx 10^{-2}$ at time~$t\approx5000$.

This first observation naturally raises a question. A naive approach would be to estimate the asymptotic speed by locally averaging the homogeneous FKPP formula, i.e.,
\begin{align}
v_{\mathrm{naive}}
&= 2\,\overline{\sqrt{1+\eta}}
\label{eq:vnaive_def}\\
&=
\begin{cases}
2-\dfrac{\delta^{2}}{12} + O(\delta^{4}), & \delta\to 0,\\[6pt]
\dfrac{2}{3}\,2^{3/2}\simeq 1.886 < 2, & \delta=1.
\end{cases}
\label{eq:vnaive_limits}
\end{align}
indicating a \emph{decrease} of the asymptotic velocity.

\begin{figure}
    \centering
    \includegraphics[width=0.7\linewidth]{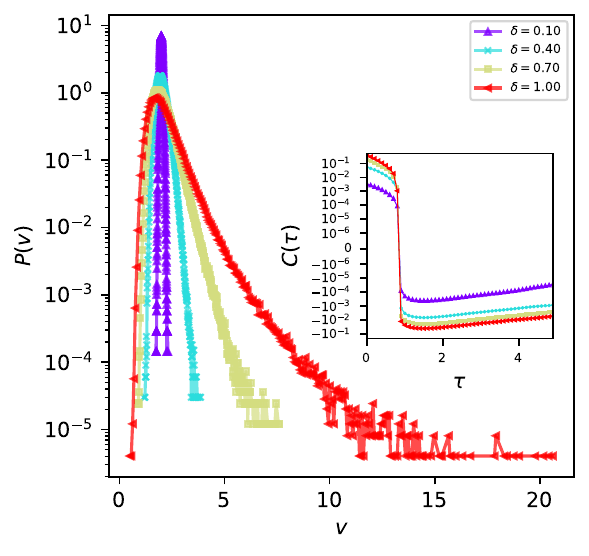}
    \caption{{\bf Instantaneous speed.} Distribution of increments $v = \frac{x(t+dt)-x(t)}{dt}$ (for $dt=0.1$) for the uniform distribution with increasing range~$\delta$, shown in the legend. Inset : autocorrelation function of the instanteaneous speed at short times, in symlog-lin scale.}
    \label{fig:inst_speeds}
\end{figure}

We now examine the increases of the front's position $x(t)$ during short time intervals of duration $dt$ (in the inset of Fig.~\ref{fig:micstudy}, and observed irregular growth w.r.t a smooth progression at speed~$\approx2$). 
In the ordered case $\eta(x)=0$, the front propagates asymptotically at speed $2$, with well-known logarithmic corrections~\cite{Bramson1983}:
\begin{equation}
    x(t) = 2t - \frac{3}{2} \log t + \mathcal{O}(1),
\end{equation}
so that the histogram of its instantaneous speeds $v = \tfrac{dx}{dt}$ is essentially a sharp peak at $v=2$. In Fig.~\ref{fig:inst_speeds}, we report the histogram of measured instantaneous speeds $v$ (with sampling interval $dt=0.1$) for solutions of Eq.~\ref{eq:fkpp} where $\eta(x)$ is uniformly distributed. 
The scale and shape of the distributions appear to scale with~$\delta$. More precisely, the standard deviation of the speed distribution, $\sigma_v$, behaves as
\begin{align}
    \sigma_v \approx (1.069 \pm 0.011)\,\sigma,
\end{align}
while its skewness scales as 
\begin{align}
    \gamma_v \sim \sigma^{1.17 \pm 0.10}    
\end{align}
The quenched reaction field induces both decreases and increases of the instantaneous speed, but in an asymmetric manner---as quantified by the skewness---despite the statistical translation invariance of the reaction field. 

\section{Front shape}

\begin{figure}
    \centering
    \includegraphics[width=0.9\linewidth]{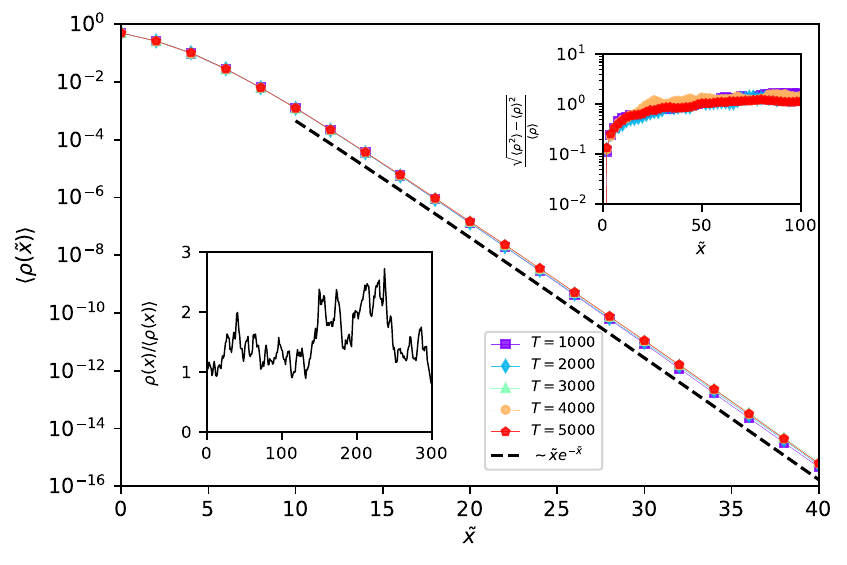}
    \caption{\textbf{Ahead of the front.} Shape of average tail $\langle \rho(\tilde{x}) \rangle$ in lin-log scale, for increasing times. Dashed line : $\tilde{x} e^{-\tilde{x}}$. Top-right corner inset: relative dispersion $\sqrt{\langle \rho^2(\tilde{x}) \rangle - \langle \rho(\tilde{x})\rangle^2}/\langle  \rho(\tilde{x})\rangle$ in function of $\tilde{x}$. Lower left corner inset : relative fluctuations $\frac{\rho(x)}{\langle \rho(x) \rangle}$ of a realization.}
    \label{fig:geometry}
\end{figure}

In the homogeneous case, the speed of the pulled front is determined by its leading-edge and particularly how it decreases. If the initial condition is dominated by $\exp(-\gamma_c x)$, the selected speed is the minimal speed (here $v_0=2$, $\gamma_c=1$)~\cite{aw2, VANSAARLOOS_2003}. Here, we examine the decay of the tail of solutions of Eq.~\ref{eq:fkpp}. In order to compare fronts shapes, we introduce the moving frame $\tilde{x}=x-x(t)$ such that $\rho(\tilde{x}=0,t)=\frac{1}{2}$ and measure the average front $
\langle \rho (\tilde{x} ) \rangle$ and its fluctuations $\sqrt{\langle \rho^2(\tilde{x}) \rangle - \langle \rho(\tilde{x})\rangle^2}$. In Fig.~\ref{fig:geometry}, we report their profiles. While the average profile exhibits a tail that decays as $\tilde{x}\,\exp(-\tilde{x})$, it also displays large fluctuations around the mean, of order $\langle \rho(x) \rangle$, as shown in the top-right inset. A fit of the tail of~$\langle \rho(x) \rangle$ in~$x^a \exp(-bx)$ gives~$a = 1.08\pm 0.12$ and~$b=1.04\pm0.04$; there is small temporal dependency. This coincides with the tail of the front in the homogeneous case, at least to the leading order.

Beyond ensemble fluctuations, one may also examine the typical roughness of the front tail. 
For a given realization of the tail profile $\rho(x)$, we define 
\begin{align}
r(\tilde{x}) = \frac{\rho(\tilde{x})}{\langle \rho(\tilde{x}) \rangle},
\end{align}
that can be interpreted as a static interface pinned with~$r(\tilde{x}=0)=1$. A realization can be observed in the lower-left corner inset of Fig.~\ref{fig:geometry}. Its roughness is measured as
\begin{align}
W = \sqrt{ \, \overline{r(x)^2} - \overline{r(x)}^{\,2} \,},
\end{align}
where $\overline{\cdot}$ denotes a spatial average. 
We find that
\begin{align}
\langle W \rangle_\sigma \sim \sigma^\lambda, 
\qquad 
\sqrt{\langle W^2 \rangle_\sigma - \langle W \rangle_\sigma^2} \sim \sigma^\kappa,
\end{align}
with~$\lambda \approx 0.94 \pm 0.04$ and $\kappa \approx 1.58 \pm 0.09$.

\section{Instantaneous velocity fluctuations}

\begin{figure}
    \centering
    \includegraphics[width=0.7\linewidth]{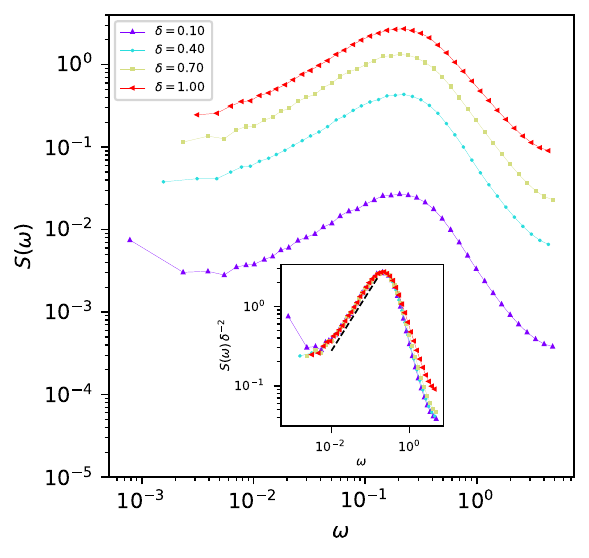}
    \caption{{\bf Spectral density for the instantaneous speed signal.} The curves correspond to various disorder intensity, for the uniform distribution. Each curve peaks around~$\omega^* \approx 2 \ . \ 10^{-1}$, preceded by a disorder-independent power-law increase~$S(\omega) \sim \omega^{0.78}$, and followed by a power-law decrease depending of the disorder intensity. Inset : collapse of the left part of the curve of the spectral density upon rescaling by~$\delta^2$. Dashed line : $\omega^{0.78}$.}
    \label{fig:spectrum}
\end{figure}

The temporal fluctuations of the front velocity are characterized by the power spectrum 
$S(\omega)=\langle|\tilde v(\omega)|^2\rangle$. 
For all disorder amplitudes $\sigma$, the spectrum exhibits a maximum at a characteristic frequency $\omega^*$ that is independent of $\sigma$ (Fig.~\ref{fig:spectrum}). 
This frequency defines a disorder-independent crossover between collective front dynamics and disorder-dominated local relaxation. 
For the uniform distribution (Fig.~\ref{fig:spectrum}), the rescaled spectra collapse for $\omega<\omega^*$, yielding
\begin{equation}
S(\omega)=\delta^2 F(\omega/\omega^*).
\end{equation}
The low-frequency behavior is universal, $F(x)\sim x^{\alpha}$ with $\alpha\simeq0.78$, independent of disorder strength, indicating that long-time velocity correlations are governed by intrinsic collective dynamics. 
In contrast, for $\omega>\omega^*$, the spectrum deviates from the Lorentzian $\sim\omega^{-2}$ and follows 
$S(\omega)\sim\omega^{\beta}$ with a disorder-dependent exponent $\beta\in[-1.75,-1.3]$, reflecting non-universal short-time fluctuations induced by quenched disorder.

We also investigate the autocorrelation of the instantaneous speed
\begin{equation}
    C(\tau) = \langle \underline{v(t) v(t+\tau)} \rangle_\sigma - \langle\underline{v(t)}\rangle_\sigma^2
\end{equation}
where we take successively the average $\underline{\cdot}$ over time and $\langle.\rangle_\sigma$ over the ensemble with disorder of standard deviation $\sigma$. We display $C(\tau)$ at short times in the inset of Fig.~\ref{fig:inst_speeds} -- long-times correlations lose their meaning given that the signal~$v(t)$ is not stationary: it has logarithmic corrections and long timescales are irrelevant. 
For all the intensities of the noise, the correlations change of sign around $\tau \approx 1$ and reach a minimum at $\tau^* \approx 1.5$ with $C(\tau^*) \sim \sigma^2$. This early sign change indicates the presence of short-time anti-correlations in the velocity fluctuations. Notably, the cancellation time $\tau \approx 1$ is essentially independent of the noise intensity $\sigma$, suggesting that it is controlled by an intrinsic dynamical timescale rather than by the amplitude of the disorder. Although the spectral density exhibits a maximum at a much lower frequency ($\tau \ll 1/\omega^* \approx 5$), this spectral feature does not dominate the short-time behavior of $C(\tau)$. Instead, the rapid decorrelation and early zero crossing point to a broadband and strongly damped noise, for which high-frequency components and fast negative feedback govern the temporal correlations (and therefore depend on $\sigma$).

\section{Asymptotic spreading speed}

We now investigate systematically the front position of the solutions of Eq.~\ref{eq:fkpp}, considering different disorder distributions and varying the disorder intensity~$\sigma$. Our goals are (i) to determine the ensemble-averaged front speed, denoted by $\langle v \rangle_{\sigma}$, and (ii) to characterize the fluctuations of the front dynamics around this mean value.
 
\begin{figure}
    \centering
    \includegraphics[width=0.7\linewidth]{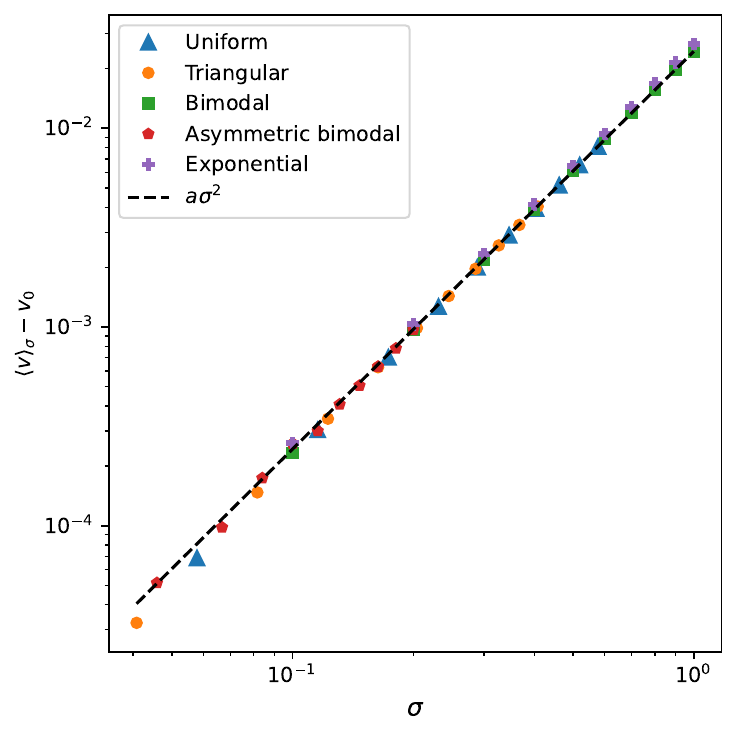}
    \caption{{\bf Quadratic speed increase.} Measured average speed increments $\langle v \rangle_\sigma - v_0$ at time $t=5000$ as a function of the disorder intensity $\sigma$, for four types of disorder: uniform, triangular, symmetric bimodal, and asymmetric bimodal. The black dashed line represents the fit in~$a \sigma^2$.}
    \label{fig:univ_speed}
\end{figure}

We compute the speed increment $\langle v \rangle_\sigma - v_0$ at time $t=5000$. 
A nonlinear regression as a function of the disorder intensity $\sigma$ yields
\begin{align} \label{eq:speed_rel}
\langle v \rangle_\sigma = v_0 + a\,\sigma^2,
\end{align}
with a fitted coefficient 
\begin{align}
a = 0.02432 \pm 0.00002
\end{align}
obtained consistently for all disorder distributions considered here (see Fig.~\ref{fig:univ_speed}).
This result (i) quantifies the acceleration of FKPP fronts induced by disorder in the reaction field, and (ii) shows the apparent universality of the coefficient $a$, independent from the microscopic details of the distribution of $\eta(x)$. 
We assume that differences arising from logarithmic corrections to the front dynamics are negligible compared to the observed asymptotic speed shift; however, this assumption is expected to break down in the limit $\sigma \to 0$.

\section{Fluctuations around the asymptotic speed}

\begin{figure}
    \centering
    \includegraphics[width=0.7\linewidth]{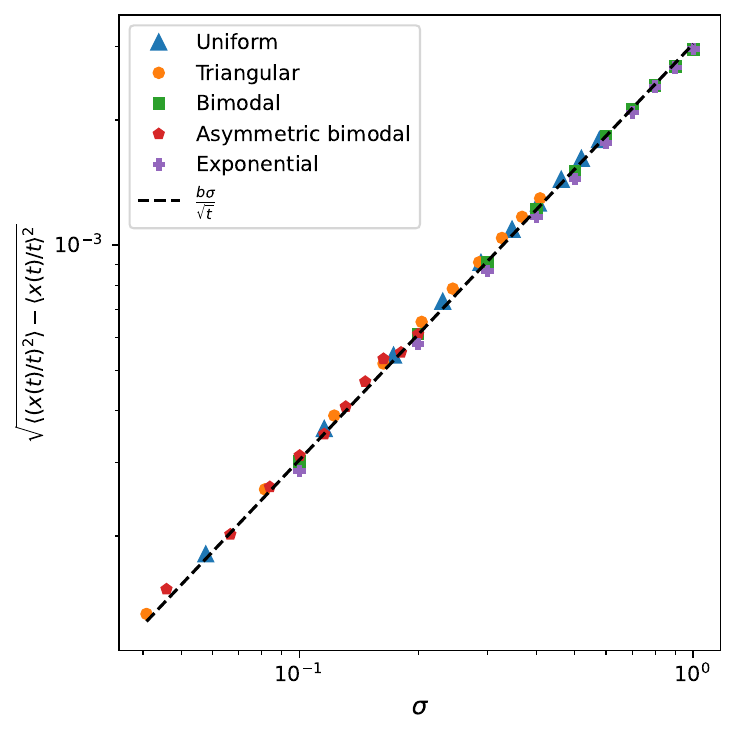}
    \caption{{\bf Diffusion coefficient.} Ensemble fluctuations of the front speed,
    $\sqrt{\langle (x(t)/t)^2\rangle - \langle x(t)/t \rangle^2}$,
    as a function of the disorder intensity $\sigma$ at time $t=5000$.
    The dashed line corresponds to the scaling $\tfrac{b}{\sqrt{t}}\sigma$.}
    \label{fig:univ_fluct}
\end{figure}

Next we compute the fluctuations of the front position as a function of the disorder intensity $\sigma$ at time $t=5000$. 
A linear regression yields
\begin{equation} 
\sqrt{\langle x(t)^2 \rangle - \langle x(t) \rangle^2}
= b\,\sigma\,\sqrt{t},
\label{eq:front_fluct}
\end{equation}
with the fitted value independent from the disorder distribution (see Fig.~\ref{fig:univ_fluct})
\begin{align}
b = 0.223 \pm 0.002
\end{align}
This scaling indicates that the front position exhibits diffusive fluctuations, with an effective diffusion coefficient
\begin{equation}
D_{\mathrm{eff}} = \frac{1}{2} b^2\,\sigma^2,
\end{equation}
which is proportional to $\sigma^2$ and independent of the specific disorder distribution.

\section{Comparison with a periodic medium}

\begin{figure}
    \centering
\includegraphics[width=1\linewidth]{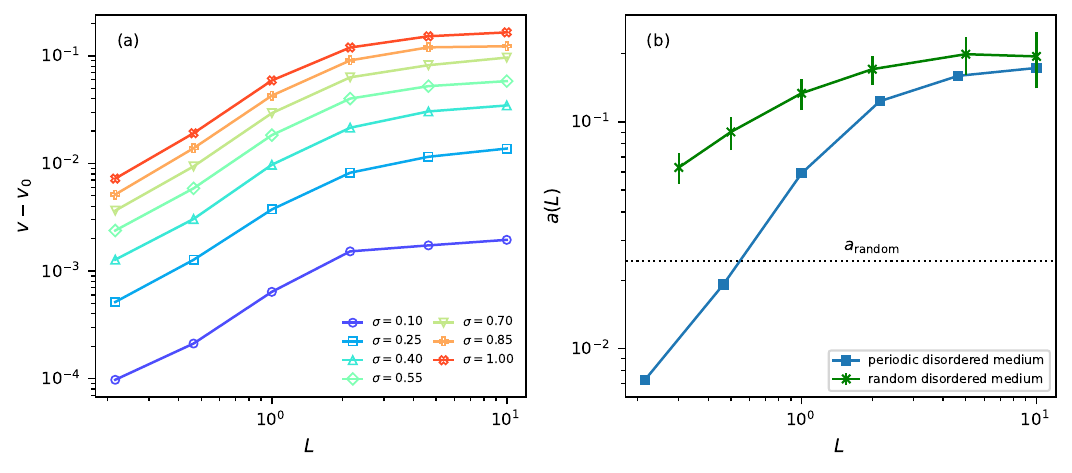}
    \caption{{\bf Deterministic periodic medium.} (a) Dependency of the speed increase~$v-v_0$ against~$L$ for various~$\sigma$. (b) Prefactor of the scaling $v_{\sigma,L} - v_0 = a(L)\sigma^2$ as a function of $L$ for a deterministic periodic medium (blue curve) and for a random disordered medium with length range $L$ (green curve). Note the monotonic increase of the function $a(L)$. The dashed horizontal line at the level $a_{\text{random}}$ represents the prefactor obtained for a fully random growth field. It intersects the curve $a(L)$ at $L_c \approx 0.5$.}
    \label{fig:periodic}
\end{figure}

We have seen that oscillations around a mean value enhance the spreading speed relative to the homogeneous case. To further probe the effect of randomness, it is instructive to compare the impact of identical heterogeneous oscillations of $K$ arranged periodically or randomly. For this purpose, we consider a periodic growth field in which $\eta$ alternates between the values $\eta = +\sigma$ and $\eta = -\sigma$ over intervals of length $L$ (i.e., with period $2L$). Its standard deviation is clearly $\sigma$, but it exhibits spatial correlations. Computing the asymptotic speed at various~$L$ and~$\sigma$, we find that it is well described (at leading order, see an exact expression in~\cite{Hamel_2011}) by
\begin{equation}
    v_{\sigma,L} -v_0\simeq a(L) \sigma^2 \,,
\end{equation}
with a similar quadratic dependency in~$\sigma$ as for random growth fields, but with a non-trivial prefactor~$a(L)$. It was conjectured by Shigesada \emph{et al.}~\cite{Shigesada_1997} that $a(L)$ increases with $L$, and this property was later established rigorously by Nadin~\cite{Nadin_2010}. Our computations shown in Fig.~\ref{fig:periodic} indeed agree with this monotonicity. They also agree with the limit of $a(L)$ that can be derived from the Theorem 2.3 from~\cite{Hamel_2011} when~$L$ goes to infinity and to~$0$.

Below a critical length~$L_c \approx 0.5$ the speed increase for deterministic  periodic media is smaller than the random one, while it exceeds this value for~$L > L_c$ (at equal disorder intensity~$\sigma$). For the sake of comparison, 
we also computed the velocity of fronts propagating in heterogeneous fields with the same constant ``high'' (and symmetrically ``low'') values over intervals of length $L$ but with the intervals arranged randomly rather than periodically.
We look at the factor representing a quadratic dependence on the disorder amplitude $\sigma$, as previously observed for pointwise random disorder and for periodic~$a = \frac{v - v_0}{\sigma^2}$. We observe that: (i) $a$ is a non-decreasing function of $L$; (ii) it is systematically larger than the corresponding prefactor for the periodic medium, indicating a stronger speed enhancement; and (iii) in the large-$L$ limit, the two prefactors converge to the same limit. This last fact would follow from the result ~\cite{Hamel_2011} once we know that the random prefactor os higher than the periodic one. 
These observations suggest the following interpretation. In a heterogeneous medium, the leading edge of the front samples both regions with reduced growth rate $\eta < \overline{\eta}$ and enhanced growth rate $\eta > \overline{\eta}$, which respectively decrease and increase the local propagation speed. Importantly, these effects are asymmetric: the acceleration induced in regions with $\eta > \overline{\eta}$ outweighs the deceleration in the complementary regions (see also Fig.~\ref{fig:inst_speeds}). As a result, the asymptotic front speed is increased on average compared to the homogeneous case. The spatial extent of these favorable regions plays a crucial role, as already emphasized in, e.g.,~\cite{Nadin_2015}. In a periodic medium, regions with enhanced growth rate have a fixed length. In contrast, in a randomly disordered medium, favorable regions occur with a broad distribution of lengths. In particular, the probability of observing a region of length $L$ over which the spatially averaged growth rate exceeds its mean obeys a large deviation principle:
\begin{align}
P\!\left( \frac{1}{L} \int_0^L \eta(x) \, dx > \overline{\eta} + \Delta \eta \right)
\sim \exp\!\left[- L \, I(\overline{\eta} + \Delta \eta)\right],
\end{align}
where $I$ is the associated rate function characterizing the disorder~\cite{Touchette_2009}. Although such regions are rare, their presence leads to a stronger enhancement of the front speed than in the periodic case, and accounts for the observed difference between random and periodic disorder.

\section{Discussion}

We studied numerically the propagation of one-dimensional Fisher--KPP fronts with steep initial conditions in quenched disordered media, characterized by a spatially quenched growth rate whose disorder amplitude is quantified by $\sigma$. The instantaneous front speed follows a skewed distribution, with its standard deviation increasing linearly with $\sigma$, and its skewness superlinearly. The tail of the average front profile decays as $x\,\exp(-x)$, with relative fluctuations of order unity. 

By computing both the autocorrelation function and the spectral density of the instantaneous speed signal, we analyzed the signal associated with the instantaneous speed of the front. The autocorrelation function reveals the oscillatory nature of the dynamics through a change of sign at a characteristic delay $\tau \approx 1$. Moreover, the spectral density exhibits a pronounced peak at a characteristic frequency $\omega^{*}$, preceded by a power-law increase
$S(\omega)\sim \omega^{0.78}$, independently of the disorder intensity. At higher frequencies, the spectrum displays a power-law decay
$S(\omega)\sim \omega^{\nu(\sigma)}$, with an exponent $\nu(\sigma)$ ranging from approximately $-1.7$ at low $\sigma$ to $\nu(\sigma)\approx -1.3$ at larger disorder amplitudes.

Finally, we find that solutions of Eq.~\eqref{eq:fkpp} satisfy
\begin{equation}
\frac{\langle x(t) \rangle_\sigma}{t} \xrightarrow[t\to\infty]{} v_0 + a\,\sigma^2,
\end{equation}
while at long times the fluctuations of the front position scale as
\begin{equation}
\sqrt{\langle x(t)^2 \rangle_\sigma - \langle x(t) \rangle_\sigma^2}
\sim b\,\sigma\,\sqrt{t}.
\end{equation}
The coefficients $a = 0.02432 \pm 0.00002$ and $b = 0.223 \pm 0.002$ are found to be independent of the microscopic details of the disorder distribution. The speed increase appears to be stronger than the one found for periodic media. Hence, random disordered heterogeneous environments enhance the spreading speed over the ordered periodic ones. We proposed a heuristic explanation based on the rare apparition of long profitable ranges for random media, further supported by simulations with random disordered growth rate constant on ranges of growing length.

These findings call for a theoretical interpretation. It is well known that Fisher--KPP fronts are pulled and are therefore driven by the dynamics in the low-density region where $\rho \approx 0$. Linearizing Eq.~\eqref{eq:fkpp} around this unstable state yields
\begin{equation}
\partial_t \rho = \partial_{xx} \rho + (1+\eta)\,\rho \,.
\end{equation}
Introducing the Cole--Hopf transformation $\rho = \exp(h)$, one finds that the field $h(x,t)$ obeys
\begin{equation}
\label{eq:qkpz}
\partial_t h = \partial_{xx} h + (\partial_x h)^2 + 1 + \eta(x),
\end{equation}
which formally resembles a Kardar--Parisi--Zhang (KPZ) equation (see also chapter 7 of~\cite{VANSAARLOOS_2003}) with an additional constant driving force $F=1$ and a quenched noise term $\eta(x)$.

This equation differs, however, from the standard quenched KPZ (QKPZ) formulation typically studied in the context of pinning--depinning transitions~\cite{barabasistanley}, where the disorder depends both on the spatial coordinate and on the interface height, i.e.\ $\eta=\eta(x,h)$. In the present case, the disorder is \emph{columnar}, being purely spatial and independent of $h$. As a consequence, the interface is not subject to pinning, but instead experiences spatially heterogeneous growth rates that persist along the direction of propagation (see e.g.~\cite{lopez1995,de_Queiroz_2008,Haldar_2020,Haldar_2021, Szendro_2007} for literature related to the columnar quenched KPZ equation).

Equation (29) is related to localization phenomena in disordered media~\cite{Szendro_2007}. In such systems, rare favorable
regions of the disorder shape the interface and leads to nontrivial roughening~\cite{Ramasco_2000}. Understanding how these rare-event tails affect front propagation remains an important open problem in the theory of fluctuating fronts.



Finally, we summarize a number of conjectures and open questions raised by our findings:
(1) Does the universality of the relations Eq.~\ref{eq:speed_rel} and Eq.~\ref{eq:front_fluct} extend beyond the cases considered here, and can the coefficients $a$ and $b$ be estimated analytically?
(2) Can the position of the invading front, starting from a step-function initial condition, be described with greater accuracy?
(3) How much larger is the increase in speed in random media with disorder constant over intervals of length $L$ compared to the periodic case?
(4) Is there a physical interpretation of the period $L_c$ in terms of an effective length scale for the growth rate in the limit of pointwise disorder?
(5) Can large-deviation techniques be used to establish this comparison rigorously? \\

\subsection*{Acknowledgements}
The authors are grateful to Julien Berestycki for his thoughtful suggestions. UM acknowledges the use of the computing cluster of the Mathematics department of the University of Trento.


\markboth{}{}

\bibliographystyle{apsrev4-2}

\appendix

\renewcommand{\thefigure}{A\arabic{figure}}
\setcounter{figure}{0}

\renewcommand{\thesection}{A\arabic{section}} 
\setcounter{section}{0}   

\clearpage
\newpage

\clearpage


%

\end{document}